\def\nabstar#1{\nabla\kern-0.5pt\smash{\raise 4.5pt\hbox{$\ast$}}
               \kern-4.5pt_{#1}}

\def\drvstar#1{\partial\kern-0.5pt\smash{\raise 4.5pt\hbox{$\ast$}}
               \kern-5.0pt_{#1}}

\def\newline{\relax\ifhmode\null\hfil\break\else\nonhmodeerr@\newline\fi}
\def\frac#1#2{{#1\over#2}}
\def\text#1{{\hbox{\rm #1}}}

\newcommand{\beq}{\begin{equation}}
\newcommand{\eeq}{\end{equation}}
\newcommand{\bea}{\begin{eqnarray}}
\newcommand{\eea}{\end{eqnarray}}

\def\BE{\begin{equation}}
\def\EE{\end{equation}}
\def\BA{\begin{eqnarray}}
\def\EA{\end{eqnarray}}
\def\BAN{\begin{eqnarray*}}
\def\EAN{\end{eqnarray*}}

\def\nn{\nonumber\\}

\def\det{\mbox{det}}

\def\gm5{\gamma_5}

\documentclass{ws-proc}
\begin{document}

\title{Optimal Lattice Domain-Wall Fermions with Finite
$N_s$\footnote{\uppercase{I}nvited talk given at 2002
\uppercase{I}nternational \uppercase{W}orkshop on
\uppercase{S}trong \uppercase{C}oupling \uppercase{G}auge
\uppercase{T}heories and \uppercase{E}ffective
\uppercase{F}ield \uppercase{T}heories (\uppercase{SCGT02}),
\uppercase{N}agoya,
\uppercase{J}apan, 10-13 \uppercase{D}ec 2002.}
}

\author{Ting-Wai Chiu\footnote{\uppercase{W}ork partially
supported by grants \uppercase{NSC91-2112-M002-025} and
\uppercase{NSC-40004F} of the
\uppercase{N}ational \uppercase{S}cience \uppercase{C}ouncil,
\uppercase{ROC}.}}

\address{Physics Department, National Taiwan University \\
Taipei, Taiwan 106, Taiwan\\
E-mail: twchiu@phys.ntu.edu.tw}


\maketitle

\abstracts{
I review the lattice formulations of vector-like gauge theories (e.g. QCD)
with domain-wall fermions,
and discuss how to optimize the chiral symmetry for any finite $ N_s $
(in the fifth dimension), as well as to eliminate its dependence on $a_5$.}

\section{Introduction}

A viable approach to study strongly-coupled gauge theories (e.g. QCD)
is to formulate these theories on a spacetime lattice with domain-wall
fermions.
The basic idea of domain-wall fermions (DWF)\cite{Rubakov:bb,Callan:sa}
is to use an infinite set of coupled Dirac fermion fields
$ \{ \psi_s (x), s \in (-\infty, \infty) \} $ with masses
behaving like a step function $ m(s) = m \theta(s) $ such that Weyl 
fermion states can arise as zeromodes bound to the mass defect 
at $ s = 0 $. However, if one uses a compact set of masses, then
the boundary conditions of the mass (step) function must
lead to the occurrence of both left-handed and right-handed
chiral fermion fields, i.e., a vector-like theory.
For lattice QCD with DWF\cite{Kaplan:1992bt}, in practice,
one can only use a finite number ($ N_s $) of
lattice Dirac fermion fields to set up the domain wall, thus the
chiral symmetry of the quark fields (in the massless limit) is broken.
Obviously, the discretization in the fifth dimension also introduces
the lattice spacing $ a_5 $ into the theory.
Presumably, only in the limit $ N_s \to \infty $ and $ a_5 \to 0 $, the
correct effective 4D theory with exact chiral symmetry can be recovered.
Now the relevant question is how to minimize its dependence on
$ N_s $ and $ a_5 $. Since, in general, they are independent
parameters, it may happen that even in the limit $ N_s \to \infty $,
the massless quark propagator is exactly chirally symmetric but still
has a strong dependence on $ a_5 (\ne 0) $.
If this is the case, then the limit ($ a_5 \to 0 $) has to be taken
before one can measure physical observables at any finite lattice spacing
$ a $, which of course, is difficult for any practical computations.

It turns out that the conventional DWF action with open boundary
conditions\cite{Shamir:1993zy} suffers from:
(i) The chiral symmetry of the quark propagator
is {\it not} optimal for any finite $ N_s $;
(ii) The quark determinant and propagator
are {\it sensitive} to the lattice spacing $ a_5 $,
even for very large $ N_s $.

In this talk, I present a formulation\cite{Chiu:2002ir,Chiu:2003ir}
of lattice QCD with DWF,
in which the quark propagator is $ a_5 $-invariant and has
optimal chiral symmetry for any $ N_s $ and background gauge
field.

\section{Problems of the conventional domain-wall fermions}

First, we outline the basic features of DWF on the lattice.
In general, given DWF action
$ {\bf A}_f [\psi,\bar\psi] $ with fermion fields
$ \{ \psi(x,s), \bar\psi(x,s); s=1,\cdots,N_s \} $, one can construct
the quark fields $ q(x), \bar q(x) $ from the boundary modes, and obtain
the quark propagator\footnote{Here the color and Dirac indices are suppressed.}
in a background gauge field as
\bea
\label{eq:quark_prop}
\langle q(x) \bar q(y) \rangle =
\frac{ \int [d\bar\psi][d\psi] q(x) \bar q(y) e^{-{\bf A}_f}}
     { \int [d\bar\psi][d\psi] e^{-{\bf A}_f}}
= ( D_c + m_q )_{x,y}^{-1}
\eea
where $ m_q $ is the bare quark mass,
\bea
\label{eq:Dc}
D_c = \frac{1 + \gamma_5 S}{1-\gamma_5 S} \ , \hspace{4mm}
\label{eq:S}
S = \frac{\prod_{s=1}^{N_s}(1+a_5 H_s) - \prod_{s=1}^{N_s} (1-a_5 H_s)}
         {\prod_{s=1}^{N_s}(1+a_5 H_s) + \prod_{s=1}^{N_s} (1-a_5 H_s)} \ ,
\eea
and $ \{ H_s, s=1,\cdots,N_s \} $ are Hermitian operators which
depend on $ {\bf A}_f $.

For the conventional DWF, $ H_s $ is the same for all $ s $,
and is equal to
\bea
\label{eq:H}
H = H_w ( 2 + \gamma_5 a_5 H_w )^{-1}, \hspace{4mm} H_w = \gamma_5 D_w \ ,
\eea
where $ D_w $ is the standard Wilson-Dirac operator plus a negative
parameter $ -m_0 $ ($0 \le m_0 \le 2$). In the limit $ N_s \to \infty $,
$ S $ becomes $ \mbox{sgn}(H) $,
\bea
\lim_{N_s \to \infty} S = \frac{a_5 H}{\sqrt{a_5^2 H^2}}
                        = \frac{H}{\sqrt{H^2}}  \ ,
\eea
and the quark propagator possesses exact chiral symmetry
\bea
\lim_{\stackrel{N_s \to \infty}{m_q \to 0}}
[ (D_c + m_q)^{-1} \gamma_5 + \gamma_5 (D_c + m_q)^{-1} ] = 0 \ .
\eea
However, for the conventional DWF, the quark propagator (\ref{eq:quark_prop})
still depends on $ a_5 $ through $ H $ (\ref{eq:H}).
It turns out that the effects of $ a_5 $ cannot be neglected
even for very large $ N_s $. This is the essential
difficulty encountered in lattice QCD calculations with
the conventional DWF.
So the relevant problem for lattice QCD with DWF is how
to construct a DWF action such that its $ S $ operator is
independent of $a_5$, for any $ a $, $ N_s $, and gauge background.

Another difficulty of the conventional DWF is that
it does {\it not} preserve the chiral symmetry optimally
for any finite $ N_s $\cite{Chiu:2002ir}.
In other words, its $ S $ operator
\bea
\label{eq:SH}
S(a_5 H) = \frac{(1+a_5 H)^{N_s} - (1-a_5 H)^{N_s}}
                {(1+a_5 H)^{N_s} + (1-a_5 H)^{N_s}}
         \equiv a_5 H R(a_5^2 H^2)
\eea
is {\it not} the optimal rational approximation of $ \mbox{sgn}(H) $.
The deviation of $ S(a_5 H) $ from $ \mbox{sgn}(H) $
can be measured in terms of
\bea
\label{eq:sign_error}
\sigma(S) &=& \max_{\forall {Y\ne0}}
\left| \frac{Y^{\dagger} \{ \mbox{sgn}(H)-S(a_5 H) \} Y}{Y^{\dagger} Y} \right|
\le \max_{\{\eta\}} \left| \mbox{sgn}(\eta) - S(\eta) \right| \ ,
\eea
where $ \{\eta\} $ are eigenvalues of $ a_5 H $.
Using the simple identity
\bea
\label{eq:sign_sqrt}
|\mbox{sgn}(x)-S(x)|=|1-\sqrt{x^2} R(x^2)| \ , \hspace{4mm} S(x) = x R(x^2)
\eea
which holds for any $ x \ne 0 $ and $ S(x) = x R(x^2) $, we can rewrite
(\ref{eq:sign_error}) as
\bea
\label{eq:sign_isqrt_error}
\sigma(S) \le \max_{\{\eta^2 \}}
\left| 1-\sqrt{\eta^2} R(\eta^2) \right|
\eea
where $ \{ \eta^2 \} $ are eigenvalues of $ a_5^2 H^2 $.
For the conventional DWF, the r.h.s. of
(\ref{eq:sign_isqrt_error}) is {\it not} the minimum for any given $ N_s $,
i.e., $ R(x^2) $ is {\it not} the optimal rational approximation
of $ (x^2)^{-1/2} $.
Obviously, the problem of finding the optimal rational approximation
$ S_{opt}(x) = x R_{opt}(x^2) $ of $ \mbox{sgn}(x) $ with
$ x \in [-x_{max},-x_{min}] \cup [x_{min},x_{max}] $
is equivalent to finding the optimal rational
approximation $ R_{opt}(x^2) $ of
$ (x^2)^{-1/2} $ with $ x^2 \in [x_{min}^2, x_{max}^2] $.

According to de la Vall\'{e}e-Poussin's theorem
and Chebycheff's theorem\cite{Akhiezer:1992},
the necessary and sufficient condition for an irreducible
rational polynomial
\BAN
r^{(n,m)}(x)=
\frac{ p_{n} x^{n} + p_{n-1} x^{n-1} + \cdots + p_0 }
     { q_{m} x^{m} + q_{m-1} x^{m-1} + \cdots + q_0 } \ ,
     \ ( m \ge n, \ p_i, q_i > 0 )
\EAN
to be the optimal rational polynomial of the inverse square root function
$ x^{-1/2} $, $ 0 < x_{min} \le x \le  x_{max} $ is that
$ \delta(x) \equiv 1 - \sqrt{x} \ r^{(n,m)}(x) $ has $ n + m + 2 $
alternate change of sign in the interval $ [x_{min}, x_{max}] $,
and attains its maxima and minima (all with equal magnitude), say,
\BAN
\delta(x) =  -\Delta, +\Delta, \cdots, (-1)^{n+m+2} \Delta
\EAN
at consecutive points ($ x_i, i=1,\cdots, n+m+2 $)
\BAN
x_{min} = x_1 < x_2 < \cdots < x_{n+m+2} = x_{max}\ .
\EAN
In other words, if $ r^{(n,m)} $ satisfies the above condition,
then its error
\BAN
\sigma(r^{(n,m)}) = \max_{x \in [x_{min},x_{max}]}
                    \left| 1 - \sqrt{x} \ r^{(n,m)}(x) \right|
\EAN
is the minimum among all irreducible rational polynomials of degree $ (n,m) $.

It is easy to show\cite{Chiu:2002ir} that $ R(x^2) $ (\ref{eq:SH}) of the
conventional DWF is {\it not} the optimal rational approximation for
$ (x^2)^{-1/2} $.
The optimal rational approximation for the inverse square root function
was first obtained by Zolotarev\cite{Zol:1877} in 1877, using
Jacobian elliptic functions.
A detailed discussion of Zolotarev's result can be found in
Akhiezer's two books\cite{Akhiezer:1992}.

Thus the relevant problem for lattice QCD with DWF is to construct
a DWF action such that the operator $ S $ in the quark propagator
(\ref{eq:quark_prop}) is equal to
\bea
S_{opt} = S_{opt}(H_w)
= \left\{ \begin{array}{ll}
          H_w R_Z^{(n,n)}(H_w^2) \ ,   &  N_s = 2n+1, \\
          H_w R_Z^{(n-1,n)}(H_w^2) \ , &  N_s = 2n,   \\
          \end{array} \right.
\label{eq:S_opt_RZ}
\eea
where $ R_Z(H_w^2) $ is the Zolotarev optimal rational
polynomial\cite{Zol:1877,Akhiezer:1992}
\bea
\label{eq:rz_nn}
R^{(n,n)}_Z(H_w^2) = \frac{d_0}{\lambda_{min}}
\prod_{l=1}^{n} \frac{ 1+ h_w^2/c_{2l} }{ 1+ h_w^2/c_{2l-1} } \ ,
\hspace{8mm}  h_w^2 = H_w^2/\lambda_{min}^2
\eea
and
\bea
\label{eq:rz_n1n}
R^{(n-1,n)}_Z(H_w^2) = \frac{d'_0}{\lambda_{min}}
\frac{ \prod_{l=1}^{n-1} ( 1+ h_w^2/c'_{2l} ) }
     { \prod_{l=1}^{n} ( 1+ h_w^2/c'_{2l-1} ) } \ ,
\eea
and the coefficients $ d_0 $, $ d'_0 $, $ c_l $ and $ c'_l $
are expressed in terms of elliptic functions\cite{Akhiezer:1992}
with arguments depending only on $ N_s $
and $ b = \lambda_{max}^2 / \lambda_{min}^2 $
($ \lambda_{min} $ and $ \lambda_{max} $ are the minimum and the
maximum of the eigenvalues of $ |H_w| $).

\section{The optimal domain-wall fermions}

Recently, I have constructed a new lattice DWF
action\cite{Chiu:2002ir,Chiu:2003ir} such that the quark
propagator is $a_5$-invariant and preserves the chiral symmetry
optimally for any $ N_s $ and background gauge field.
Further, its effective 4D lattice Dirac operator for the internal
fermion loops is shown to be exponentially-local
for sufficiently smooth gauge backgrounds\cite{Chiu:2002kj}.

Explicitly, the optimal lattice domain-wall fermion action
reads\footnote{In this paper, we suppress the lattice spacing $a$,
as well as the Dirac and color indices, which can be restored easily.}
\bea
\label{eq:ODWF}
{\bf A}_f &=& \sum_{s,s'=0}^{N_s+1} \sum_{x,x'}
\bar\psi(x,s)
\{ (\omega_s a_5 D_w(x,x') + \delta_{x,x'} )\delta_{s,s'} \nn
&& \hspace{1mm}   -(\delta_{x,x'}-\omega_s a_5 D_w(x,x'))
    (P_{-} \delta_{s',s+1} + P_{+} \delta_{s',s-1}) \} \psi(x',s')
\eea
with boundary conditions
\bea
\label{eq:bc1_new}
P_{+} \psi(x,-1) &=& - r m_q P_{+} \psi(x,N_s+1) \ ,
\hspace{4mm} r = \frac{1}{2m_0} \\
\label{eq:bc2_new}
P_{-} \psi(x,N_s+2) &=& -r m_q P_{-} \psi(x,0) \ ,
\eea
and the quark fields constructed from the left and right boundary modes
\bea
\label{eq:q_odwf}
q(x) = \sqrt{r} \left[ P_{-} \psi(x,0) + P_{+} \psi(x,N_s+1) \right] \ , \\
\label{eq:bar_q_odwf}
\bar q(x)= \sqrt{r} \left[ \bar\psi(x,0)P_{+}+\bar\psi(x,N_s+1)P_{-}\right] \ ,
\eea
where
the weights $ \{ \omega_s \} $ are given by
\bea
\omega_0 &=& \omega_{N_s+1} = 0 \ , \\
\omega_s a_5 &=& \frac{1}{\lambda_{min}} \sqrt{ 1 - \kappa'^2 \mbox{sn}^2
             \left( v_s ; \kappa' \right) }, \hspace{6mm} s=1,\cdots,N_s \ .
\label{eq:omega}
\eea
Here $ \mbox{sn}( v_s; \kappa' ) $ is the Jacobian elliptic function
with argument $ v_s $ (\ref{eq:vs}) and modulus
$ \kappa' = \sqrt{ 1 - 1/b } $, where
$ b = \lambda_{max}^2 / \lambda_{min}^2 $.

For the optimal DWF, $ H_s = \omega_s H_w $ in the operator
$ S $ (\ref{eq:S}), and the weights in (\ref{eq:omega}) are obtained from
the roots $ (u_s = (\omega_s a_5)^{-2}, s=1,\cdots,N_s) $ of
the equation
\bea
\label{eq:delta_Z}
\delta_Z(u) =
     \left\{ \begin{array}{ll}
 1-\sqrt{u} R_Z^{(n,n)}(u)=0 \ ,   & \ N_s=2n+1 \nn
 1-\sqrt{u} R_Z^{(n-1,n)}(u)=0 \ , & \ N_s=2n   \nn
             \end{array} \right.
\eea
such that $ S $ is equal to $ S_{opt} $ (\ref{eq:S_opt_RZ}),
the optimal rational approximation of $ \mbox{sgn}(H_w) $,
and the quark propagator (\ref{eq:quark_prop}) has the
optimal chiral symmetry for any $ N_s $ and
$ b = \lambda_{max}^2/\lambda_{min}^2 $.

The argument $ v_s $ in (\ref{eq:omega}) is
\bea
\label{eq:vs}
v_s &=& (-1)^{s-1} M \ \mbox{sn}^{-1}
   \left( \sqrt{\frac{1+3\lambda}{(1+\lambda)^3}}; \sqrt{1-\lambda^2} \right)
 + \left[ \frac{s}{2} \right] \frac{2K'}{N_s}
\eea
where
\bea
\label{eq:lambda}
\lambda =
\prod_{l=1}^{N_s}
\frac{\Theta^2 \left(\frac{2lK'}{N_s};\kappa' \right)}
     {\Theta^2 \left(\frac{(2l-1)K'}{N_s};\kappa' \right)} \ ,
\hspace{2mm}
M =
\prod_{l=1}^{[\frac{N_s}{2}]}
\frac{\mbox{sn}^2 \left(\frac{(2l-1)K'}{N_s};\kappa' \right) }
{ \mbox{sn}^2 \left(\frac{2lK'}{N_s};\kappa' \right) } \ ,
\eea
$ K' $ is the complete elliptic integral of the first kind with
modulus $ \kappa' $, and $ \Theta $ is the elliptic theta function.
From (\ref{eq:omega}), it is clear that
$ \lambda_{max}^{-1} \le \omega_s a_5 \le \lambda_{min}^{-1} $,
since $ \mbox{sn}^2(;) \le 1 $.

Obviously, the optimal domain-wall fermion action (\ref{eq:ODWF})
is invariant for any $ a_5 $,
since its dependence on $ a_5 $ is only through the product $ \omega_s a_5 $
(\ref{eq:omega}) which only depends on $ N_s $, $ \lambda_{min} $ and
$ \lambda_{max}^2/\lambda_{min}^2 $. Therefore, {\it $ a_5 $ is a
redundant parameter in the optimal domain-wall fermion action}
(\ref{eq:ODWF}).

The generating functional for $n$-point
Green's function of quark fields $ q $ and $ \bar q $
is defined as
\bea
\label{eq:ZW}
Z[J,\bar J]
= \frac{ \int [dU][d\bar\psi][d\psi][d\bar\phi][d\phi]
         e^{ -{\bf A}_g -{\bf A}_f -{\bf A}_{pf}
   + \sum_x \left\{ \bar J(x)q(x) + \bar q(x)J(x) \right\}} }
       {\int [dU][d\bar\psi][d\psi][d\bar\phi][d\phi]
         e^{-{\bf A}_g-{\bf A}_f-{\bf A}_{pf}} }
\eea
where $ {\bf A}_g $ is the gauge field action, $ {\bf A}_f $
is the domain-wall fermion action, $ {\bf A}_{pf} $ is the
corresponding pseudofermion action with $ m_q=2m_0 $,
and $ \bar J $ and $ J $ are
the Grassman sources of $ q $ and $ \bar q $ respectively.
The purpose of introducing the pseudofermion fields (which carry
all attributes of the fermion fields but obey the Bose statistics)
is to provide the denominator $ (1+rD_c)^{-1} $ in the effective 4D
lattice Dirac operator
\bea
\label{eq:Dm}
D(m_q) = (D_c + m_q)(1 + r D_c )^{-1},
\hspace{2mm}
D_c = 2 m_0 \left(\frac{1+\gamma_5 S_{opt}}{1-\gamma_5 S_{opt}} \right)
\eea
for internal fermion loops such that $ D(m_q) $ is
exponentially-local\cite{Chiu:2002kj} (for any $ N_s $ and $ m_q $),
and its fermion determinant ratio
(in the limit $ N_s \to \infty $ and $ a \to 0 $)
is equal to the corresponding
ratio of $ \det[\gamma_\mu (\partial_\mu + i A_\mu) + m_q] $.

Evaluating the integrations over $ \{\psi,\bar\psi \} $ and
$ \{ \phi, \bar\phi \} $ in (\ref{eq:ZW}),
one obtains\cite{Chiu:2003ir}
\bea
\label{eq:ZW_odwf}
Z[J,\bar J] =
\frac{ \int [dU] e^{-{\bf A}_g[U] } \det[D(m_q)]
       e^{ \bar J  ( D_c + m_q )^{-1} J }  }
     { \int [dU] e^{-{\bf A}_g[U]}  \det[D(m_q)] }
\eea

Then any $n$-point Green's function can be obtained by
differentiating $ Z[J,\bar J] $ with respect to $ J $ and $ \bar J $
successively. In particular, the quark propagator is
\BAN
\langle q(x) \bar q(y) \rangle =
\left.
\frac{\delta^2 Z[J,\bar J]}{\delta \bar J(x) \delta J(y)}
\right|_{J=\bar J=0}
\hspace{-2mm}
= \frac{ \int [dU] e^{-{\bf A}_g} \det [D(m_q)]
          (D_c + m_q)_{x,y}^{-1} }
       { \int [dU] e^{-{\bf A}_g} \det [D(m_q)] }
\EAN
which, in a background gauge field, becomes
\bea
\label{eq:quark_prog}
  \langle q(x) \bar q(y) \rangle
= ( D_c + m_q )_{x,y}^{-1} \ ,
\eea
and it goes to
$ [ \gamma_\mu ( \partial_\mu + i A_{\mu} ) + m_q ]^{-1} $ in the
limit $ N_s \to \infty $ and $ a \to 0 $. For any gauge background
and $ N_s $, $ D_c $ has the optimal chiral symmetry since the error
$ \sigma(S_{opt}) $ is the minimum, and
$ \sigma(S_{opt}) \le (1-\lambda)/(1+\lambda) \simeq A(b)e^{-c(b)N_s} $,
where $ \lambda $ is defined in (\ref{eq:lambda}), $ A(b) $
and $ c(b) $ can be estimated as
$ A(b) \simeq 4.06(1) b^{-0.0091(1)} \mbox{ln}(b)^{0.0042(3)} $,
$ c(b) \simeq 4.27(45) \mbox{ln}(b)^{-0.746(5)} $,
$ b = \lambda_{max}^2/\lambda_{min}^2 $.

Finally, we note that $ D(0) $ (\ref{eq:Dm}) ($N_s \to \infty$)
times a factor $ r = 1/2m_0 $ is exactly equal to the overlap
Dirac operator\cite{Neuberger:1998fp}
$ D_o = [ 1 + D_w (D_w^{\dagger} D_w)^{-1/2} ]/2 $.
This implies that $ D $ is topologically-proper
(i.e., with the correct index and axial anomaly),
similar to the case of overlap Dirac operator.
For any finite $ N_s $, $ r $ times $ D(0) $ (\ref{eq:Dm})
is exactly equal to the overlap Dirac operator with $ (H_w^2)^{-1/2} $
approximated by Zolotarev optimal rational polynomial,
(\ref{eq:rz_nn}) ($N_s$ = odd), or (\ref{eq:rz_n1n}) ($ N_s $ = even).
Further discussions can be found in Ref. [6].

\section*{Acknowledgments}

I am grateful to Koichi Yamawaki and Yoshio Kikukawa
for inviting me to this interesting workshop
and to all organizers of SCGT02 for their kind hospitality.


\begin{thebibliography}{0}


\bibitem{Rubakov:bb}
V.~A.~Rubakov and M.~E.~Shaposhnikov,
Phys.\ Lett.\ B {\bf 125}, 136 (1983).


\bibitem{Callan:sa}
C.~G.~Callan and J.~A.~Harvey,
Nucl.\ Phys.\ B {\bf 250}, 427 (1985).


\bibitem{Kaplan:1992bt}
D.~B.~Kaplan,
Phys.\ Lett.\ B {\bf 288}, 342 (1992)


\bibitem{Shamir:1993zy}
Y.~Shamir,
Nucl.\ Phys.\ B {\bf 406}, 90 (1993)

\bibitem{Chiu:2002ir}
T.~W.~Chiu,
Phys.\ Rev.\ Lett.\  {\bf 90}, 071601 (2003)

\bibitem{Chiu:2003ir}
T.~W.~Chiu,
hep-lat/0303008.

\bibitem{Akhiezer:1992}
N.~I.~Akhiezer,
{\it Theory of approximation} (Dover, New York, 1992);
%
{\it Elements of the theory of elliptic functions},
Translations of Mathematical Monographs, 79,
(American Mathematical Society, Providence, RI. 1990).

\bibitem{Zol:1877}
E.~I.~Zolotarev,
Zap. Imp. Akad. Nauk. St. Petersburg, 30 (1877), no. 5; reprinted in his
Collected works, Vol. 2, Izdat, Akad. Nauk SSSR, Moscow, 1932, p. 1-59.



\bibitem{Chiu:2002kj}
T.~W.~Chiu,
Phys.\ Lett.\ B {\bf 552}, 97 (2003)


\bibitem{Neuberger:1998fp}
H.~Neuberger,
Phys.\ Lett.\ B {\bf 417}, 141 (1998);
%
%
R.~Narayanan and H.~Neuberger,
Nucl.\ Phys.\ B {\bf 443}, 305 (1995)



\end{thebibliography}
\end{document}